\documentclass[a4paper, 10 pt, conference]{ieeeconf} 

\IEEEoverridecommandlockouts   
\usepackage{graphics} 
\usepackage{graphicx} 
\usepackage{epstopdf}
\usepackage[dvipsnames]{xcolor}
\usepackage{epsfig} 
\usepackage{amsmath} 
\usepackage{amssymb}  
\usepackage{algorithm,algpseudocode}
\usepackage{subfigure}
\usepackage{multirow}
\usepackage{booktabs}
\usepackage{caption}
\usepackage{array}
\usepackage{bbm}

\newcommand{\D}{\mathcal{D}}

\newcommand{\T}{\mathcal{T}}

\newcommand{\M}{\mathcal{M}}

\newcommand{\R}{\mathbb{R}}

\newtheorem{theorem}{Theorem}

\newtheorem{proposition}[theorem]{Proposition}
\newtheorem{problem}{Problem}

\title{\LARGE \bf
	Learning Markov models of fading channels in wireless control networks: a regression trees based approach
}

\author{Luis Felipe Florenzan Reyes$^1$, Francesco Smarra$^1$, Yuriy Zacchia Lun$^2$, and Alessandro D'Innocenzo$^1$
	\thanks{The research leading to these results has received funding from the Italian Government under CIPE resolution n.135 (Dec. 21, 2012), project INnovating City Planning through Information and Communication Technologies (INCIPICT).}
	\thanks{$^1$ Department of Information Engineering, Computer Science and Mathematics,
		University of L'Aquila, 67100 AQ, Italy.{\tt \scriptsize luisfelipe.florenzanreyes@graduate.univaq.it, francesco.smarra@univaq.it, alessandro.dinnocenzo@univaq.it}
		$^2$ IMT School for Advanced Studies Lucca, 55100 LU, Italy.
		{\tt\scriptsize	yuriy.zacchialun@imtlucca.it}}%
}

\begin{document}

	\maketitle
	
	\begin{abstract}
		Finite-state Markov models are widely used for modeling wireless channels affected by a variety of non-idealities, ranging from shadowing to interference. In an industrial environment, the derivation of a Markov model based on the wireless communication physics can be prohibitive as it requires a complete knowledge of both the communication dynamics parameters and of the disturbances/interferers.
		In this work, a novel methodology is proposed to learn a Markov model of a fading channel via historical data of the signal-to-interference-plus-noise-ratio (SINR).
		Such methodology can be used to derive a Markov jump model of a wireless control network, and thus to design a stochastic optimal controller that takes into account the interdependence between the plant and the wireless channel dynamics. The proposed method is validated by comparing its prediction accuracy and control performance with those of a stationary finite-state Markov chain derived assuming perfect knowledge of the physical channel model and parameters of a WirelessHART point-to-point communication based on the IEEE-802.15.4 standard.
	\end{abstract}
	
	
	\section{Introduction}
	Wireless networked control systems (WNCSs) are composed of spatially distributed sensors, actuators, and controllers communicating through wireless networks \cite{Park2018}. Despite their success in industrial monitoring applications, existing wireless sensor-actuator network technologies face significant challenges in supporting control systems due to their lack of real-time performance and dynamic wireless conditions in industrial plants \cite{Lu2016}. A key challenge in WNCSs design is the channel modeling in an industrial environment because of its inherent complexity \cite{Lu2016} \cite{Alhen2019}. These communication channels are frequently subject to time-varying fading and interference, which may lead to packet losses. 
	
\begin{figure}[h!]
		\begin{center}
			\centerline{\includegraphics[width=0.65\columnwidth]{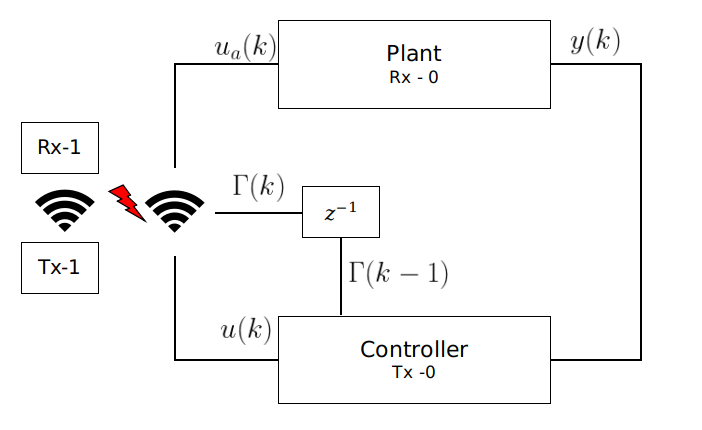}}
			\caption{Wireless networked control system}
			\label{fig:WNCS}
		\end{center}
	\end{figure}

	From the automatic control perspective, an example of WNCS consists of a nonlinear process with intermittent control packets due to the lossy communication channel described by the following equations:
	
	\begin{equation}\label{eq:mjs}
	\begin{cases}
	y\left( k+1 \right)=f \left(y(k), u_a(k)\right), k \in \mathbb{N} \\
	u_a(k) = \nu(k)u(k) \\
	y(0) = y_0 \in \mathbb{R}^{n_y}
	\end{cases}
	\end{equation}
	where $y(k) \in \mathbb{R}^{n_y}$ is the output of the system and $u(k)\in \mathbb{R}^{n_u}$ is the input signal. $\left\{\nu(k)\right\}_{k\in\mathbb{N}}$ is a discrete-time Boolean process modeling the packet delivery of the control signals: if the packet is correctly delivered then $\nu(k)=1$, otherwise if it is lost then the actuator does nothing $\nu(k)=0$. The control packet error probability (PEP) $\mathbb{P}(\nu(k)=0)$, depends on the signal-to-interference-plus-noise-ratio (SINR) of the communication $\Gamma(k)$, i.e., 
	\begin{equation}\label{eq:Bern}
	\mathbb{P}(\nu(k)=0) = g\left( \Gamma (k)\right),
	\end{equation}
	where $g:\mathbb{R} \to \left[0,1\right]$ is a deterministic function defined by the communication standard.  We assume full state observation with no measurement noise, and no observation packet loss.
	In this work, we consider a channel model based on WirelessHART \cite{Chen2010}, a on-the-market wireless communication standard specifically designed for process automation. 
	We assume that the value of the SINR $\Gamma(k)$ is measurable and is sent to the controller via an acknowledgment (ACK).
	However, this ACK is available only after the current decision on the control input to apply has been made and sent through the link, since the actual success of the transmission is not known in advance.
	Hence, at each time $k$ the controller receives measurements of $y(k)$ and $\Gamma(k-1)$, as depicted in Fig.~\ref{fig:WNCS}.
	
	\vspace{-20pt}
	
	Since $\Gamma(k)$ is represented by a generic stochastic process, the obtained model may  not be 
	computationally tractable when it is derived for wireless communications in industrial environment (see Sec.~\ref{sec:channel}), especially if the objective is to apply optimal control algorithms (e.g. model predictive control -- MPC). For this reason, a preliminary investigation of channel model abstraction is fundamental to design a controller.
	In the WNCS literature the packet dropouts have been modeled either as
	stochastic or deterministic phenomena. For what concerns
	stochastic models, a vast amount of research assumes memoryless packet drops, so that dropouts are realizations of a Bernoulli process.
	In \cite{Schenato2007}, the packet delivery process is modeled as a Bernoulli random process, then an optimal controller is derived. In \cite{Lun2019} a Markov chain model is used to derive an accurate abstraction of the WirelessHART channel, and it is proven that such model allows to characterize the stability of a WNCS for the scenarios where the simple Bernoulli-like channel models (which cannot model packet bursts) fail. 
	
	Markov models are a powerful tool for modeling stochastic random processes. They are general enough to model with high accuracy a large variety of processes and are relatively simple, allowing us to compute analytically many important parameters of the process which are very difficult to calculate for other models \cite{Turin1998}. Hidden Markov models (\cite{Bilmes1998b} \cite{Lawrance1989}) have been exploited to learn channels models in \cite{Alhen2019} and \cite{Salamatian2001}. In \cite{Sadghi2008}, the authors expose the benefit of exploiting finite-state Markov chains 
	to model the behavior of wireless fading channels. In \cite{Lun2020}, the authors derive a Markov chain that estimates the packet error probability (PEP) of an industrial wireless protocol, and then propose an optimal stochastic controller for linear systems.   
	
	Inspired by the challenges in \cite{Lun2020} related to the derivation of a physics-based Markov chain abstraction of the wireless channel, the main contribution of this paper is a novel data-driven methodology, based on regression trees \cite{Breiman2017classification}, to identify such Markov chain abstraction. More precisely, we propose a novel methodology to model the PEP $\nu(k)$ on the basis of a Markov chain $\theta(k)$: each state of $\theta(k)$ is associated to a partition of $\mathbb{R}^{{n_y}+1}$ consisting of rectangular sets $\{R_i\}_{i=1}^{\ell}$, each representing the range of possible values of $\Gamma(k) \in \mathbb R$ and $y(k) \in \mathbb{R}^{{n_y}}$ at time $k$. We construct the transition probability matrix (TPM) of $\theta(k)$ as follows:
	\begin{align}\label{eqMCidentification}\small
	p(j \mid i) 	&\doteq \mathbb{P}(\theta(k+1)=j \mid \theta(k)=i)\notag \\
	&= \mathbb{P}(\theta(k+1)=j \mid (\Gamma(k), y(k)) \in R_i)\notag \\
	&= h(\Gamma(k), y(k)),
	\end{align}
	where identifying from historical data the function $h:\mathbb{R}^{{n_y}+1} \to \left[0,1\right]$, which depends on the current measurements $\Gamma(k)$ and $y(k)$, is the objective of this paper. Note that, given any two states $i,j$ of the Markov chain, $p(j\mid i)$ also depends on the plant output: indeed, $y(k)$ may for example determine the distances between the transmitter, the receiver and the interferer, and, as illustrated in Sec.~\ref{sec:channel}, this strongly affects the dynamics of $\Gamma(k)$. Finally, the PEP can be easily computed using 
	\eqref{eq:Bern}, which in this paper is based on a point-to-point WirelessHART communication based on the IEEE 802.15.4 standard.
	Nevertheless, we remark that our data-driven methodology is independent on the transmission technology and can be replicated for any communication protocol if the corresponding wireless channel 
	can be effectively modeled by a Markov chain.   
	
	In summary, in an industrial environment the derivation of a Markov model based on the wireless communication physics can be prohibitive as it requires a complete knowledge of both the communication dynamics parameters and of the disturbances/interferers. Our methodology has 3 main advantages: (1) it only exploits historical data, hence does not require any a-priori knowledge of the system and channel parameters. Moreover, most of physics based approaches cannot handle time-varying parameters, as the computational complexity of the obtained abstraction would be intractable, while with our methodology we consider a dependency between the parameters of the communication system and the dynamics of the plant which is still tractable in terms of computational complexity; (2) as a byproduct of leveraging our techniques in \cite{SmarraNAHS2020}, and beyond the main contribution of identifying the Markov chain abstraction of $\Gamma(k)$, we also construct a switching auto-regressive exogenous (SARX) model for the nonlinear plant $f$ in \eqref{eq:mjs}, i.e. our methodology does not require any linearity assumption on the plant's dynamics; (3) the obtained identified models, both for $\Gamma(k)$ and $f$, can be combined obtaining a Markov jump system, which can be directly used to setup a classical MPC problem that can be solved very efficiently, i.e. using quadratic programming (QP).
	
	
	\section{Channel modeling} \label{sec:channel}
	In this section we first describe the channel model under consideration, then we present the Markovian model for the wireless link developed in \cite{Lun2020}, and finally we introduce an analytical model that will be useful to experiment the data-driven approach we propose in this paper.
	
	
	\subsubsection{WirelessHART physical model} 
	We analyze the industrial environment described in Fig. \ref{fig:WNCS}, wherein the wireless communication is affected by interference.
	We study a point to point transmission based on the WirelessHART protocol, i.e. the IEEE 802.15.4-2006 defined in \cite{2006}, interfered by another WirelessHART transmission.  
	The proposed model considers the effect of path loss, the shadow fading, and the residual power fluctuations left by the power control. The effect of multipath fading is supposed to be compensated by the aforementioned power control. We denote with $i\!=\!0$ the reference Transmitter-Receiver (Tx-Rx) link, and with $i\!=\!1$ the link between the interferer Tx and the reference Rx.
	
	The effect of the path loss model is defined in \cite{2006}. For a system with bandwidth $W\!=\!2.4$GHz the path loss coefficient of the link between transmitter $i$ and the reference receiver (i.e., a mobile plant in our case) is $\alpha_i(k) \!=\! 10^{-\frac{\varsigma(d_i(k))}{10}}$, where
	\begin{gather}\label{eq:path}
	\varsigma(d_i(k)) = 
	\begin{cases}
	40.2+20\log_{10}(d_i(k)), \text{ if } d_i(k) \leq 8, \\
	58.5 + 33 \log_{10}(\frac{d_i(k)}{8}), \text{ otherwise;} 
	\end{cases}	
	\end{gather}
	and $d_i(k)$ is the length of the link $i$ at time instant $k$, i.e.~a distance in meters that may e.g.~depend on the position of the plant (see the inverted pendulum on a cart in Sec.~\ref{sec:NR}).
	
	The shadow fading is modeled following \cite{Goldsmith2005} by assuming a log-normal model for each link $i$, which introduces a multiplicative factor $e^{\beta_i(k)}$, where $\beta_i(k)$ is a zero-mean Gaussian process with variance $\sigma_{\!\beta_i}^{2}$ and auto-covariance function $c_{\beta_i}(\tau)$, with $\tau$ being a time lapse between two consecutive (time-driven) control packets. We remark that $c_{\beta_i}(\tau)$ may also depend on the state of the plant (e.g. the speed of a cart) thus exhibiting a time-varying behavior.
	
	For each link $i$ the residual power control error (PCE) is also modeled as a log-normal process, $e^{\xi_i(k)}$, where $\xi_i(k)$ is a zero-mean Gaussian process with variance $\sigma_{e_i}^2$ and auto-covariance $c_{\xi_i}(\tau)$.
	
	By considering the characteristics of the offset quadrature phase-shift keying (OQPSK) with direct-sequence spread spectrum (DSSS) modulation, as specified in \cite{2006}, 
	we can derive the power value of the SINR, $\gamma(k)$, at time $k$ 
	as in \cite{Lun2020}:  
	\begin{equation}\label{eq:sinr}
	\gamma(k) = 	\sqrt{\frac{P_0(k)\alpha_0^2(k)e^{\beta_0(k)+\xi_0(k)}}{\frac{N_0}{4}+\frac{8}{3G}P_1(k)\alpha_1^2(k)e^{\beta_1(k)+\xi_1(k)}}},
	\end{equation}
	where $P_0$ and $P_1$ are respectively the reference user transmission power and the interferer transmission power, $N_0$ is the noise spectral density, and $G=WT_s$ is the processing gain, with $T_s$ being the symbol time.
	In the rest of the work we denote as $\Gamma(k)=10\log_{10}(\gamma(k))$ the SINR in decibel. 
	
	In WirelessHART the forward error correction is not implemented, thus even one erroneous bit leads to a corrupted WirelessHART data packet.
	For this reason, the packet error probability $R_p$ is related to the SINR as follows
	\begin{align} \label{eq:Rp}
	R_b(\gamma(k))&=\frac{1}{30} \sum_{\iota=2}^{16}(-1)^{\iota} \binom{16}{\iota}\exp\left(20\cdot \gamma(k)\frac{1-\iota}{\iota}\right)\nonumber \\
	R_p(\gamma(k))&=1-\left(1-R_b\left(\gamma\left(k\right)\right)\right)^{l_f},
	\end{align}
	where $l_f$ is the number of bits in a frame, and $R_b$ is the 
	bit error rate computed according to \cite{2006}. 
	It is worth remarking that the distance Tx-Rx $d_i(k)$ influences the SINR $\gamma(k)$ that, in turn, influences the packet error rate. This will be useful for the discussion of the simulation results in Sec.~\ref{sec:NR}. 
	
	The main issue with the above channel model is that, despite the ability to describe the SINR, in the majority of the control applications equation \eqref{eq:sinr} is intractable, both in the continuous-time and discrete-time form. For this reason, in \cite{Lun2020}, a Markov chain model to derive a discrete-time abstraction of \eqref{eq:sinr} has been proposed. We recall such technique in the following subsection as a comparison for the method we propose in this paper. 
	
	
	\subsubsection{Finite-state Markov chain}\label{ssec:FSMC}
	It is straightforward to see that \eqref{eq:sinr} can be expressed as a weighted linear combination of correlated log-normal processes: hence there is no exact explicit closed form expression of its distribution.
	We can use the moment matching technique \cite{fischione2007} to approximate the probability distribution of \eqref{eq:sinr} by a log-normal process, thus presenting 
	$\Gamma(k)$ as a Gaussian process with mean $\mu_{\Gamma}(k)$, variance $\sigma_{\Gamma}^2(k)$ and auto-covariance $c_{\Gamma}(\tau)$, as detailed in \cite{Lun2020}. Clearly, 
	due to the dependence of the SINR on the state of a mobile plant through, e.g. the parameters $d_i(k)$ (which define the path loss coefficients $\alpha_i(k)$ via \eqref{eq:path}), 
	the moment matching approximation should be done for each relevant value of the aforementioned parameters.
	In the rest of this subsection we will focus on a Gaussian process $\hat{\Gamma}(k)$, which is a moment matching approximation of the SINR with arbitrary values of the parameters in \eqref{eq:sinr} corresponding to any given state of the plant, i.e. for $\alpha_i(k)\!=\!\hat{\alpha}_i$, $\beta_i(k)=\hat{\beta}_i(k)$, and $P_i(k)\!=\!\hat{P}_i$ in \eqref{eq:sinr}, with $i\!=\!0,1$,  we have that $\hat{\Gamma}(k)\approx 10\log_{10}(\gamma(k))$.
	
	At this point, to obtain a finite-state Markov channel abstraction, the first step is to divide the range of $\hat{\Gamma}(k)$ into several consecutive regions, each associated with a certain representative PEP. Specifically, a region $r$ of the values of the SINR is mapped into a state $\boldmath{s_r}$ of the related Markov chain, and it is delimited by two thresholds $\zeta_r$ and $\zeta_{r+1}$ belonging to the set of extended reals. These SINR thresholds are determined by the chosen partitioning method \cite{Sadghi2008}. 
	In this paper, we rely on a well-known equiprobable partitioning, 
	where the thresholds are selected in such a way that the steady state probabilities of being in any state are equal.
	
	Then, the steady state probability $\boldsymbol{p_r}$ of a state $\boldmath{s_r}$ is defined as the probability that the value of $\hat{\Gamma}(k)$ is between the two thresholds of the region, and it is given by
	\begin{equation}
	\boldsymbol{p_r} = \int_{\zeta_r}^{\zeta_{r+1}} \phi_{\mathcal{N}}(\zeta;\mu_{\hat{\Gamma}},\sigma_{\hat{\Gamma}}^2)d\zeta,
	\end{equation}
	where $\mu_{\hat{\Gamma}}$ and $\sigma_{\hat{\Gamma}}^2$ are respectively the mean and variance of $\hat{\Gamma}(k)$, and $\phi_{\mathcal{N}}(\zeta;\mu,\sigma^2)$ is the probability density function (PDF) of a Gaussian random variable (GRV) $\mathcal{N}(\mu,\sigma^2)$. 
	
	The Packet Delivery (PDP) associated to the same state within the respective region is given by 
	\begin{equation}
	1-\nu_M^{(r)} = \frac{1}{\boldsymbol{p_{r}}} \int_{\zeta_r}^{\zeta_{r+1}}R_p (10^{\frac{\zeta}{10}})\phi_{\mathcal{N}}(\zeta;\mu_{\hat{\Gamma}},\sigma_{\hat{\Gamma}}^2)d\zeta.
	\end{equation}
	
	Finally, the channel state transition probabilities are derived from integrating the joint PDF 	of the SINR $\hat{\Gamma}(k)$ over two consecutive packet transmissions and over the desired regions, $r$ and $q$, as
	\begin{equation*}
	p(q\mid r)\!=\!\frac{1}{\boldsymbol{p_r}}\!\!
	\int_{\zeta_{r}}^{\zeta_{r+1}}\!\!\!\!\int_{\zeta_{q}}^{\zeta_{q+1}}\!\!\!\!\!
	\varphi_{\mathcal{N}}\!\left(\varsigma_{k-1},\varsigma_{k};\mu_{\hat{\Gamma}},\sigma_{\hat{\Gamma}}^2,c_{\hat{\Gamma}}(\tau)\!\right)\!d\varsigma_{k-1}d\varsigma_{k},
	\end{equation*}
	where $\varphi_{\mathcal{N}}(\varsigma_{k-1},\varsigma_{k};\mu_{\hat{\Gamma}},\sigma_{\hat{\Gamma}}^2,c_{\hat{\Gamma}}(\tau))$ is the two-dimensional PDF of the Gaussian process {$\hat{\Gamma}(k)$}, as detailed in \cite{Lun2020}.
	
	However, one issue related to the above technique is that in real (industrial) cases some of the channel parameters required for the above modeling are time-varying (due to their dependence on the state of the plant) and often only partially known.
	For these reasons, we propose a new approach to model the SINR, and thus the PEP. In this respect, a promising direction is the exploitation of historical data of the communication channel to identify a model via system identification and machine learning techniques. 
	
	
	\subsubsection{Auto-regressive model for channel simulation}\label{sec:AR}
	
	The application of data-driven methodologies requires the existence of a dataset containing trajectories of the SINR: in this paper we run Monte Carlo simulations of the wireless transmission model \eqref{eq:sinr} and then apply our techniques to such trajectories.
	To this aim, in this section we illustrate the approach presented in \cite{Baddour2005}, where the study of auto-regressive stochastic models for computer simulation of fading channels is addressed, to derive discrete-time trajectories of the process described in equation \eqref{eq:sinr}. 
	In particular, let us consider a discrete-time Gaussian process $\left\{z(k)\right\}_{k\in \mathbb{N}}$ with auto-correlation function (ACF) $R_{zz}(n)$.
	We can derive an auto-regressive (AR) model of the following form that is able to generate trajectories of the process:
	\begin{equation}
	z(k) = -\sum\nolimits_{n=1}^{p}a_nz(k-n)+w(k),
	\end{equation}
	where $w(k)$ is a zero-mean white Gaussian noise process.
	The AR model parameters consist of coefficients $\{a_1,\ldots, a_p\}$ and variance $\sigma_p^2$ of the driving noise $w(k)$.
	To estimate the coefficients $a_j$, $j = 1,\ldots,p$, once the ACF $R_{zz}$ is fixed from $\beta_i$ and $\xi_i$, $i = 0,1$, the relationship between $R_{zz}$ and $a_j$ is given as follows \cite{Kay1988}:
	\begin{equation}\label{eq:Rzz} \small
	R_{zz}(n) =
	\begin{cases} 
	-\sum_{m=1}^{p}a_mR_{zz}(n-m), & n \geq 1 \\ 
	-\sum_{m=1}^{p}a_mR_{zz}(-m) + \sigma_p^2, & k=0 
	\end{cases}
	\end{equation}
	%
	%
	Finally, coefficients $a_j$ can be determined solving the set of Yule-Walker equations that can be easily derived from \eqref{eq:Rzz} (we refer the reader to \cite{Kay1988} for further details).
	
	The generated AR process has the following auto-correlation function:
	\begin{equation}
	\hat{R}_{zz}(n) =
	\begin{cases} 
	R_{zz}(n), & 0\leq n \leq p \\ 
	-\sum_{m=1}^{p}a_m \hat{R}_{zz}[n-m], & n>p 
	\end{cases}
	\end{equation}
	The simulated process has the attractive property that its sampled ACF perfectly matches the desired sequence of ACF up to lag $p$.
	Therefore, since we know the ACF of both the residual power control error $\xi_i(k)$ and the shadowing correlation $\beta_i(k)$ we can exploit the AR model to generate sequences of $\xi_i(k)$ and $\beta_i(k)$ and then apply \eqref{eq:sinr} to obtain the discrete-time trajectories of $\Gamma(k)$.
	
	\section{The Classification And Regression Tree (CART) algorithm}\label{ssec:BackgroundRT}
	
	The aim of this section is to provide a short description of the classification and regression trees (CART) algorithm \cite{Breiman2017classification}, in order to provide the basic notions to present our method in Sec.~\ref{sec:RTMC} that identifies a Markov chain wireless channel abstraction.

	In a supervised framework, we consider a predictor dataset $\mathcal{P} = \{\lambda(k)\}_{k=1}^{D}$ and a response dataset $\mathcal{R}=\{\rho(k)\}_{k=1}^{D}$ of $D$ samples each, where $\rho(k) \in \mathbb{R}$ is called response variable and $\lambda(k) \in \mathbb{R}^n$ is called predictor variable. The final goal of CART is to identify a function $\T$ to estimate $\hat \rho(k)=\T(\lambda(k))$.
	
	In the specific case of the CART algorithm \cite{Breiman2017classification}, the dataset is partitioned into a set of hyper-rectangles $R_1,\ldots,R_\ell$, corresponding to the $\ell$ leaves of the tree. Then, $\hat \rho(k)$ is estimated in each leaf $\tau_i$ using a constant $c_{\tau_i}$ given by the average of the samples in the partition. 
	Without any loss of generality we restrict our attention to recursive binary partition.
	Due to space limitations, we only briefly recall the partitioning algorithm of CART, and refer the reader to \cite{Breiman2017classification} for more details.
	
	The CART algorithm creates the partition using a greedy algorithm to optimize the split variables and split points: starting with the whole dataset, consider a split variable $j$ over the $n$ available and a split point $s$, and define the pair of half-planes as $R_L(j,s) =\{\lambda(k)\mid \lambda_j(k) < s\}$ and $R_R(j,s) = \{\lambda(k) \mid \lambda_j(k) \geq s\}$. Then, CART solves the following optimization problem to find the optimal $j$ and $s$
	\begin{equation}\label{eq:minCART} \scriptsize
	\min_{j,s} \left[\min_{c_L} \sum_{\lambda(k) \in R_L(j,s)} (\rho(k)-c_L)^2 + \min_{c_R} \sum_{\lambda(k) \in R_R(j,s)} (\rho(k)-c_R)^2\right],
	\end{equation}
	\normalsize
	and for any choice of $j$ and $s$ the inner minimization is solved by $c_L  = \text{ave}(\rho(k) \mid \lambda(k) \in R_L(j,s))$ and $c_R  = \text{ave}(\rho(k) \mid \lambda(k) \in R_R(j,s))$, where $\text{ave}(\cdot)$ is the arithmetic mean of the output samples.
	In other words, the optimal $j^*$ and $s^*$ minimize the sum of the quadratic prediction errors of the left and right partitions induced by the split variable and split point. 
	
	For each splitting variable, the determination of the split point $s$ can  be done very quickly and hence, by scanning through all of the inputs, the determination of the best pair $(j, s)$ is feasible.
	Once the best split is found the dataset is partitioned into the two resulting regions, then the splitting procedure is repeated on each of the two regions.
	The process is repeated on all of the resulting regions until a stopping criterion is applied, e.g. tree size is a tuning parameter chosen to avoid overfitting and variance phenomena.
	
	In the rest of this work we denote with $\T$ the regression tree, with $\tau_i$ the $i^{th}$ leaf of $\T$, with $|\T|$ the number of leaves of $\T$, with $|\tau_i|$ the number of samples in $\tau_i$, with $c_{\tau_i} = \text{ave}(\rho(k)| \lambda(k) \in \tau_i)$ the prediction of leaf $\tau_i$ and, with a slight abuse of notation, with $\T(\lambda)$ the prediction of the regression tree, i.e. 
	\begin{equation}\label{eq:RTpred}\small
	\T (\lambda) =  \sum_{\tau_i \in \T} \sum_{ \lambda(k) \in \tau_i} \frac{\rho(k)}{|\tau_i|}  I \left\{\lambda \in \tau_i\right\} =  \sum_{\tau_i \in \T} c_{\tau_i} I \left\{\lambda \in \tau_i\right\},
	\end{equation}
	where $I \left\{\lambda \in  \tau_i \right\}$ is the indicator function, which is equal to 1 if $\lambda \in \tau_i$ and 0 otherwise. 
	
	
	\section{Switching ARX Identification}\label{ssec:BackgroundSARX}
	
	As discussed above, in this paper we leverage the techniques in \cite{SmarraNAHS2020} to construct a switching auto-regressive exogenous (SARX) model for the nonlinear plant $f$ in \eqref{eq:mjs} that can be directly used to setup a MPC problem, as will be done in Sec.~\ref{sec:NR}. In particular, starting from a dataset $\D=\{(y(k),u(k))\}_{k=1}^{D}$ of $D$ samples collected from the measurements of a physical system, respectively consisting of outputs $y(k)\in\R^{n_y}$ and inputs $u(k)\in\R^{n_u}$, we will derive for each $j=0,\ldots,N-1$ a model as follows:
	
	\footnotesize
	\begin{align}\label{eqIdentifiedModelN}
	\nonumber x(k+j+1) = A_{\sigma_j(x(k))} x(k+j)+ B_{\sigma_j(x(k))} u(k+j) + F_{\sigma_j(x(k))},
	\end{align}
	\normalsize
	
	\noindent with $\sigma_j:\R^{n_x}\rightarrow\M\subset\mathbb{N}$ the switching signal, $x(k)\doteq\left[y^\top(k)\ \cdots\ y^\top(k-\delta_y)\ u^\top(k-1)\ \cdots\ u^\top(k-\delta_u)\right]^\top\in\R^{n_x}$ the state consisting of the regressive terms of the inputs and the outputs, $\delta_y,\delta_u\geq 0$, and $n_x = (\delta_y + 1)n_y + n_u \delta_u$.

	\section{Markov model based on regression trees}\label{sec:RTMC}
	
	In industrial environments, the derivation of Markov models based on the physics of wireless communication can be prohibitive as it requires complete knowledge of both the communication dynamics parameters and the disturbances/interferers. Furthermore, the presence of time-varying parameters can increase the complexity of the obtained model. We propose a novel methodology to derive a Markov model of the PDP based on the WNCS measurements. The approach exploits the transmissions' historical data to deal with the circumstances illustrated above. In particular, we handle the case wherein there is a dependency between the communication system's and the plant measurable outputs. 
	
	The main idea is to derive a Markovian model $\left\{\theta(k) \in \Theta \right\}_{k \in \mathbb{N}}$ abstracting the SINR stochastic process $\Gamma(k)$, with TPM $P$ as in \eqref{eqMCidentification}, and associate to each state $i\in \Theta$ a PDP. The learning procedure consists of three steps: (1) we grow a regression tree $\T$ with predictor variables the current SINR and the plant measurements, i.e. $\left(\Gamma(k), x(k) \right)$, and with response variable the SINR at the next time step, i.e. $\Gamma(k+1)$. Since the leaves of $\T$ form a partition of the predictor space, we define the state-space of $\theta(k)$ associating to each element of the partition $\{R_{\tau_i}\}_{i=1}^{\ell_\T}$ a state of the Markov chain, i.e. $\Theta \doteq \left\{ i \in \mathbb{N}: \tau_i \in \T \right\}$; (2) we grow a regression tree $\Pi$ with predictor variables the current SINR, i.e. $\Gamma(k)$, and with response variable the prediction of the SINR at the next time step obtained using the regression tree $\T$, i.e. $\T(\Gamma(k),x(k))$. We will combine $\Pi$ and $\T$ to identify the TPM of $\left\{\theta(k) \right\}_{k \in \mathbb{N}}$; (3) we associate to each state of $\left\{\theta(k) \right\}_{k \in \mathbb{N}}$ a PDP.
	
	
	\subsubsection{Step 1} We grow a regression tree $\T$ using as predictor dataset $\left\{\Gamma(k), x(k) \right\}_{k=1}^{D} = \mathcal{P}_\T \subset \mathbb{R}^{n_x+1}$, consisting of the current SINR and the plant measurable outputs, and as response dataset $\left\{ \Gamma(k+1) \right\}_{k=1}^{D} = \mathcal{R}_\T \subset \mathbb{R}$, consisting of the SINR at the next time step. Let $\T(\left( \Gamma(k), x(k)\right))$, with $\T: \mathcal{P}_\T \to \mathcal{R}_\T$, be the prediction of $\T$ given the current measurements. 
	As mentioned above, we associate to each element of the partition $\{R_{\tau_i}\}_{i=1}^{\ell_\T}$ a state of the Markov chain $\left\{\theta(k) \in \Theta \right\}_{k \in \mathbb{N}}$, i.e. $\Theta \doteq \left\{ i \in \mathbb{N}: \tau_i \in \T \right\}$. Note that $\forall \left(\Gamma(k), x(k)\right) \in \mathbb{R}^{n_x+1}, \exists! \ \tau_i \in \T: \left(\Gamma(k), x(k)\right) \in R_{\tau_i}$, i.e. the current measurement $\left(\Gamma(k), x(k)\right)$ is deterministically associated to one, and only one, discrete state of $\theta(k)$. 
	
	As each leaf $\tau_i$ contains a random subset of the SINR sequence, we assume independent and identically distributed samples belonging to the same partition element. We fit a GRV $G_{\tau_i} \sim \mathcal{N}(\mu_i,\sigma_i^2)$ to model the current level of SINR in each leaf $\tau_i \in \T$ of the tree:
	\begin{gather}\small
	\begin{aligned}
	&\mu_{i} =\frac{1}{|\tau_i|}\sum_{ \left( \Gamma(k), x(k)\right)  \in R_{\tau_i}} \Gamma(k), \\
	&\sigma_{i}^2 = \frac{1}{|\tau_i|-1} \sum_{ \left( \Gamma(k), x(k)\right)  \in R_{\tau_i}} (\Gamma(k)-\mu_{i})^2.
	\end{aligned}
	\end{gather}
	In Sec.~\ref{sec:pep} we will exploit $\left\{G_{\tau_i} \right\}_{\tau_i \in \T}$ to estimate the PEP in each state of $\left\{\theta(k)\right\}_{k\in \mathbb{N}}$. 
	
	Following the same reasoning, we fit a GRV $G_{\tau_i}^+ \sim \mathcal{N}(\mu_{i^+},\sigma_{i^+}^2)$ to model the next-step level of SINR for each leaf $\tau_i \in \T$ of the tree:
	\begin{gather}\small\label{eq:cp}
	\begin{aligned}
	\mu_{i^+} &=\frac{1}{|\tau_i|}\sum_{ \left( \Gamma(k), x(k)\right) \in R_{\tau_i}} \Gamma(k+1), \\
	\sigma_{i^+}^2 &= \frac{1}{|\tau_i|-1} \sum_{ \left( \Gamma(k), x(k)\right) \in R_{\tau_i}} (\Gamma(k+1) -	\mu_{i^+})^2.
	\end{aligned}
	\end{gather}
	In Sec.~\ref{sec:TPM} we will exploit $\left\{G_{\tau_i}^+ \right\}_{\tau_i \in \T}$ to identify the TPM of $\left\{\theta(k)\right\}_{k\in \mathbb{N}}$.
	
	We remark that the TPM of $\left\{\theta(k)\right\}_{k\in \mathbb{N}}$ could be identified using only $\T$, based on the number of samples that at time $k$ stay in a leaf $\tau_i\in \T$, and at time $k+1$ jump to a leaf $\tau_j \in \T$, i.e. $p(j \mid i) = \tilde{n}(\tau_i,\tau_j) \cdot |\tau_i|^{-1}$ where
	
	\small 
	\begin{align}\label{eq:naive}
	\tilde{n}(i,j) \doteq &|\left\{ \left( \Gamma(k_\epsilon), x(k_\epsilon)\right) \in \mathcal{P}_\T : \right. \\ 
	&  \left. \left( \Gamma(k_\epsilon), x(k_\epsilon)\right) \in R_{\tau_i}, \left( \Gamma(k_\epsilon+1), x(k_\epsilon+1)\right) \in R_{\tau_j} \right\}|.\notag
	\end{align}
	\normalsize
	The lack of this approach is that the tree $\T$ partitions the dataset to minimize the prediction error of a deterministic estimate of $\Gamma(k+1)$, instead of minimizing the prediction error with respect to the estimate $\mathbb{E} \left[\T(\Gamma(k),x(k)) \mid \Gamma(k) \right]$. We overcome this issue in the next section growing an additional regression tree $\Pi$, and combining $\T$ and $\Pi$ to identify the TPM to minimize the error with respect to $\mathbb{E} \left[\T(\Gamma(k),x(k)) \mid \Gamma(k) \right]$.
	
	
	\subsubsection{Step 2}\label{sec:TPM}
	
	The idea is to define a partition $\left\{ R_{\pi_r}\right\}_r^{|\Pi|}$ of $\mathbb{R}$ minimizing the prediction error with respect to the estimate $\mathbb{E} \left[\T(\Gamma(k),x(k)) \mid \Gamma(k) \right]$ , and exploit the Markov property to split the identification process of the TPM of $\left\{\theta(k) \right\}_{k\in \mathbb{N}}$ in two steps:
	\begin{gather} \small \label{eq:tpm}
	\begin{aligned}
	&\mathbb{P}(\theta(k+1)=j \mid \theta(k)=i) \\& \doteq \mathbb{P}((\Gamma(k+1), x(k+1)) \in R_{\tau_j} \mid (\Gamma(k),x(k)) \in R_{\tau_i}) \\
	&=\sum_{\pi_r \in \Pi} \{ \mathbb{P}((\Gamma(k+1), x(k+1)) \in R_{\tau_j} \mid \Gamma(k+1) \in R_{\pi_r}) \\ &\mathbb{P}(\Gamma(k+1) \in R_{\pi_r} \mid (\Gamma(k),x(k)) \in R_{\tau_i}) \}.
	\end{aligned}
	\end{gather}
	
	\begin{figure}[h!]
		\begin{center}
			\centerline{\includegraphics[width=0.9\columnwidth]{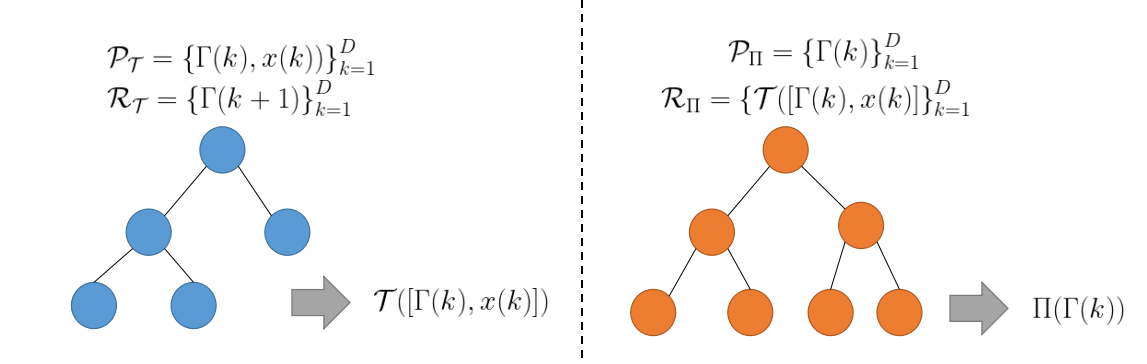}}
			\caption{TPM identification via regression trees}
			\label{figtrees}
		\end{center}
	\end{figure}
	
	To estimate $\mathbb{P}((\Gamma(k+1), x(k+1)) \in R_{\tau_j} \mid \Gamma(k+1) \in R_{\pi_r})$ we define a new predictor dataset $\{\Gamma(k) \}_{k=1}^{D} = \mathcal{P}_\Pi \subset \mathbb{R}$ and a new  response dataset $\{\T(\Gamma(k),x(k)) \}_{k=1}^{D} = \mathcal{R}_\Pi\subset \mathbb{R}$. Then, we derive a new regression tree $\Pi$ applying the CART algorithm to identify the function $\Pi: \mathcal{P}_\Pi \to \mathcal{R}_\Pi$. Let $ \Pi(\Gamma(k))$ be the optimal estimation given by the tree $\Pi$.
	Define 
	\begin{gather}\label{eq:tp} \scriptsize
	\begin{aligned}
	\mathbb{P}((\Gamma(k+1), x(k+1)) \in R_{\tau_j} \mid \Gamma(k+1) \in R_{\pi_r}) \doteq p(\tau_j\mid\pi_r) = \frac{n(\pi_r,\tau_j)}{|\pi_r|}
	\end{aligned}
	\end{gather}
	where $n(\pi_r,\tau_j)$ is the number of samples that belong both the leaf $\pi_r \in \Pi$ and the leaf $\tau_j \in \T$, i.e 
	\scriptsize
	\begin{equation}\label{eq:n}
	n(\pi_r,\tau_j) \doteq |\{( \Gamma(k_\epsilon), x(k_\epsilon)) \in \mathcal{P}_\T : \Gamma(k_\epsilon) \in R_{\pi_r}, ( \Gamma(k_\epsilon), x(k_\epsilon)) \in R_{\tau_j} \}|.
	\end{equation}
	\normalsize
	Notice that in contrast to the definition in equation \eqref{eq:naive}, equation \eqref{eq:n} does not involve the time evolution. For this reason, eq. \eqref{eq:n} is useful to estimate the probabilities $\mathbb{P}((\Gamma(k+1), x(k+1)) \in R_{\tau_j} \mid \Gamma(k+1) \in R_{\pi_r})$.
	
	\begin{proposition}\label{prop:p}
		Let us define the transition probabilities $p(\tau_j\mid\pi_r)$ as in Equation \eqref{eq:tp}, then the CART algorithm creates the partition induced by $\left\{ R_{\pi_r} \right\}_{\pi_r \in \Pi}$ optimally estimating the conditional expectation of the prediction in each leaf $\pi_r\in\Pi$, i.e.  $c_{\pi_r} =\mathbb{E} \left[\T(\Gamma(k),x(k)) \mid \Gamma(k) \in R_{\pi_r} \right], \forall \pi_r \in \Pi$.
	\end{proposition}
	
	\scriptsize
	\begin{proof}
		\begin{align}
		&c_{\pi_r} = \sum_{\Gamma(k_\epsilon) \in R_{\pi_r}} \frac{\T(\Gamma(k_\epsilon), x(k_\epsilon))}{|\pi_r|}\\
		&= \sum_{\Gamma(k_\epsilon) \in R_{\pi_r}} \frac{1}{|\pi_r|}  \sum_{\substack{\left(\Gamma(k_\epsilon'), x(k_\epsilon') \right)\in R_{\tau_j} \\ \left(\Gamma(k_\epsilon), x(k_\epsilon) \right)  \in R_{\tau_j} }}  \frac{\Gamma(k_\epsilon'+1)}{|\tau_j|}\\
		&= \sum_{\substack{\Gamma(k_\epsilon) \in R_{\pi_r}\\ \left(\Gamma(k_\epsilon), x(k_\epsilon) \right)  \in R_{\tau_j} }} \frac{1}{|\pi_r|}  \sum_{\substack{\left(\Gamma(k_\epsilon'), x(k_\epsilon') \right)\in R_{\tau_j} }}  \frac{\Gamma(k_\epsilon'+1)}{|\tau_j|}\\
		&= \sum_{\substack{\Gamma(k_\epsilon) \in R_{\pi_r}\\ \left(\Gamma(k_\epsilon), x(k_\epsilon) \right)  \in R_{\tau_j} }} \frac{c_{\tau_j}}{|\pi_r|}=\sum_{\tau_j \in \T} \frac{c_{\tau_j} \cdot n(\pi_r, \tau_j)}{|\pi_r|}  =\sum_{\tau_j \in \T} p(\tau_j \mid \pi_r)c_{\tau_j} \\
		&\simeq \sum_{\tau_j \in \T} p(\tau_j \mid \pi_r) \mathbb{E}\left[\T(\Gamma(k),x(k)) \mid \left(\Gamma(k), x(k) \right) \in R_{\tau_j}  \right] \label{eq:proof1}\\ 
		&= \mathbb{E}\left[\T(\Gamma(k),x(k)) \mid \Gamma(k) \in R_{\pi_r} \right]
		\end{align}	
	\end{proof}
	\normalsize
	In \eqref{eq:proof1} we assume that the dataset consists of independently drawn observations, and that the number of samples in each region of the tree is large enough to neglect the Standard
	Error of the sample Mean (SEM): as a consequence, the expectation can be assumed approximately equal to the sample mean. In conclusion, running the CART algorithm on our extended dataset derives transition probabilities that minimize the square of the error between the samples of the dataset $\left\{\mathcal{P}_\T , \mathcal{R}_\T \right\}$ and the corresponding conditional expectation of the predictive model of $\T$.
	
	To estimate $\mathbb{P}(\Gamma(k+1) \in R_{\pi_r} \mid (\Gamma(k),x(k)) \in R_{\tau_i})=p(\pi_r\mid\tau_i)$ we exploit the set of GRVs defined in Equation \eqref{eq:cp} and the partition induced by the tree $\Pi$: 
	\begin{equation}\label{eq:tp2}
	p(\pi_r\mid\tau_i) = \mathbb{P} \left(  G_{\tau_i}^+ \in R_{\pi_r}  \right) =\int_{R_{\pi_r}} \phi_{\mathcal{N}}(\zeta;\mu_{i^+},\sigma_{i^+}^2) d\zeta
	\end{equation}
	where $\phi_{\mathcal{N}}(\zeta;\mu,\sigma^2)$ is the PDF of the GRV $\mathcal{N}(\mu,\sigma^2)$.  
	The TPM $P = [p(j\mid i)]_{i,j}\in \mathbb{R}^{|\T|\times |\T|}$ is defined by $p(j\mid i) = \mathbb{P}(\theta(k+1)=j\mid\theta(k)=i)$ $ \forall \tau_i,\tau_j \in \T$, as in Equation \eqref{eq:tpm}.
	
	
	\subsubsection{Step 3}\label{sec:pep}
	The variable of interest in control applications is the PDP. The IEEE-802.15.4 provides the estimation of the PER given the SINR, see Equation \eqref{eq:Rp}. Let $\boldsymbol{\nu} = (\nu_1, \ldots, \nu_{|\T|})$ associate a PDP for each channel operating mode as follows:  
	\begin{equation}\label{eq:pep} \scriptsize
	1-\nu_i = E[R_p(10^{G_{\tau_i}/10}) \mid \theta(k) = i] = \int_{\mathbb{R}} R_p(10^{\zeta/10}) \phi_{\mathcal{N}}(\zeta;\mu_{c_i},\sigma_{c_i}^2)d\zeta.
	\end{equation}
	\normalsize
	
	Algorithms \ref{algPEPIde} and \ref{algPDPPre} summarize respectively the learning methodology and the procedure to estimate the packet delivery probability (PDP) over a time horizon of length $N$. 
	\begin{algorithm}[ht!]
		\small
		\caption{Learning algorithm (off-line)}
		\label{algPEPIde}
		\begin{algorithmic}[1]
			\State \textsc{Input: $\left\{\Gamma(k), x(k) \right\}_{k=1}^{D}$, Output: $\T$, $P$, $\boldsymbol{\nu}$}
			\Function{Learn PEP}{}
			\State \textsc{Define $\mathcal{P}_\T = \left\{\Gamma(k), x(k) \right\}_{k=1}^{D} , \mathcal{R}_\T= \left\{ \Gamma(k+1) \right\}_{k=1}^{D} $;}
			\State \textsc{Build $\T$ using $\mathcal{P}_\T $  to predict $\mathcal{R}_\T $;}
			\State \textsc{Define $\mathcal{P}_\Pi =\left\{\Gamma(k)\right\}_{k=1}^{D} , \mathcal{R}_\Pi =  \left\{\Pi(\left(\Gamma(k), x(k )\right)) \right\}_{k=1}^{D}$;}
			\State \textsc{Build $\Pi$ using $\mathcal{P}_\Pi $  to predict $\mathcal{R}_\Pi $;} 
			\State \textsc{Build $P$ based on \eqref{eq:tpm};}
			\State \textsc{Compute $\boldsymbol{\nu}$ using \eqref{eq:pep};}
			\State \Return $\T$, $P$, $\boldsymbol{\nu}$
			\EndFunction
		\end{algorithmic}
		\normalsize
	\end{algorithm}
	\begin{algorithm}[ht!]
		\small
		\caption{PDP estimation (run-time)}
		\label{algPDPPre}
		\begin{algorithmic}[1]
			\State \textsc{Input: $\T$, $P$, $\boldsymbol{\nu}$, $\left( \Gamma(k), x(k) \right)$, $N$, Output: $\boldsymbol{\hat\nu}$ }
			\Function{Predict PDP}{}
			\State Let $i \in \mathbb N$ be the integer such that $\left(\Gamma(k), x(k) \right) \in \tau_i$;
			\ForAll{$j=1,\ldots,N$}
			\State $\boldsymbol{\hat\nu}(j) = \boldsymbol{\nu}P^j(i,:)$, where $P^j(i,:)$ is the $i$-th row of $P^j$.
			\EndFor
			\State \Return $\boldsymbol{\hat\nu}$
			\EndFunction
		\end{algorithmic}
		\normalsize
	\end{algorithm}
	
	
	\section{Case study} \label{sec:NR}
	We consider a WNCS consisting of an inverted pendulum on a cart remotely controlled over a WirelessHART link as illustrated in Sec.~\ref{sec:channel}. In the numerical simulations, we model the plant using the nonlinear discrete-time model of the inverted pendulum and we model the packet delivery process as in equation \eqref{eq:Rp} through the SINR trajectories obtained from the AR model in Sec.~\ref{sec:AR}. We compare two implementations of Stochastic Model Predictive Control (S-MPC) \cite{Bernardini2011}: one based on the physics based channel model of Sec.~\ref{ssec:FSMC} as in \cite{Lun2020}, and one based on the data-driven channel model of Sec.~\ref{sec:RTMC}.
	\begin{figure}[h!]
		\begin{center}
			\centerline{\includegraphics[width=0.65\columnwidth]{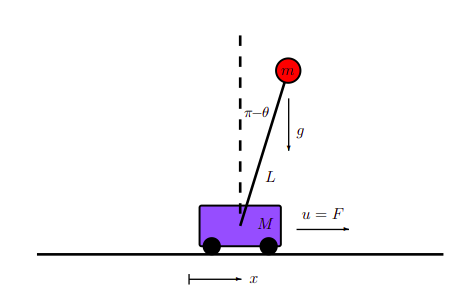}}
			\caption{Inverted pendulum on cart.}
			\label{figIPC}
		\end{center}
	\end{figure}
	
	\subsubsection{Plant model}
	We consider the following nonlinear discrete-time model $y(k+1) = y(k) + f(x(k),u_a(k))T_u$, where $y \in \R^4$, $u \in \R$ and 
	
	\begin{equation} \small
	\begin{aligned} 
	f_1(y,u) &= y_2 \\ 
	f_2(y,u) &= \frac{mL^2}{D} \left[ -mg\cos(y_3)\sin(y_3)+ mLy_4^2\sin(y_3) \right] \\ &\qquad + \frac{mL^2}{D} \left(u-\delta y_2\right)
	\\f_3(y,u) &= y_4 \\
	f_4(y,u) &= \frac{mL}{D}\left[ (m+M)g \sin(y_3)-\cos(y_3)  mLy_4^2\sin(y_3)\right]\\ &\qquad - \frac{mL\cos(y_3)}{D}\left(u-\delta y_2\right) \\
	D &= mL^2\left[ M+m\left( 1- \cos(y_3)^2\right) \right],
	\end{aligned}
	\end{equation}
	where $y_1$ is the cart position, $y_2$ is the velocity, $y_3$ is the pendulum angle, $y_4$ is the angular velocity, $m$ is the pendulum mass, $M$ is the cart mass, $L$ is the length of the pendulum arm, $g$ is the gravitational acceleration, $\delta$ is a friction damping on the cart, $u$ is a control force applied to the cart and $T_u$ is the sampling time.
	
	We analyze a case wherein the model's dynamics influence the channel behavior. More in detail, we consider that a time-varying distance between the plant and the controller, $d_0$ in equation \eqref{eq:path}, depends on the cart's position, $y_1$, i.e. $d_0(k) = y_1(k) + \bar{d}_0$,  $ \bar{d}_0 \in \mathbb{R}^+$. The other channel parameters are assumed constants. As consequence, the intermittent control packets can be modeled as follows: 
	\begin{gather}
	\begin{aligned}
	u_a(k) = \nu(k)u(k), \quad \nu(k) \sim B(1, 1-R_b(\gamma(k)) )
	\end{aligned}	
	\end{gather}
	where $u(k)$ is the input computed by the controller, $B(n,p)$ is the Bernoullian distribution with a time-varying parameter.
	Notice that $y_1(k)$ influences $\gamma(k)$ via equation $\eqref{eq:path}$.

	\subsubsection{Numerical results}
	
	The data-driven model derived in Sec.~\ref{ssec:BackgroundSARX} can be used to formalize the following:
	\begin{problem}\label{pbMPC-SA} \textit{Stochastic Model Predictive Control}
		\begin{equation*}
		\small
		\begin{aligned}
		& \underset{u_k}{\mathrm{min}} & & \mathbb{E} \left[ e_{k+N}^\top Q e_{k+N} + \sum_{j=0}^{N-1}e_{k+j+1}^\top Q e_{k+j+1} + u^\top_{k+j} R u_{k+j} \right] \\
		& \mathrm{s.t.\ }            & &  x_{k+j+1}  =   A'_{i_j}x_{k} + \sum_{\alpha = 0}^{j}{B'_{i_j,\alpha}\hat{\nu}_{k+\alpha}} u_{k+\alpha} + F'_{i_j},\\       
		&                                   & &  u_{k+j}   \in \mathcal{U}, \mathbb{E} \left[x_{k+j+1} \right] \in \mathcal{O},\\ 
		&									& &\mathbb{E} \left[x_{k+N} \right]   \in \mathcal{O}_N, x_{k}      =   x(k), j=0\ldots,N-1 ,  			\\
		\end{aligned}
		\end{equation*}
	\end{problem}
	
\noindent where $e_k = x_k -x_k^* $ is the difference between the current state of the plant and the target, $\hat{\nu}_{k+\alpha}$ is the PDP prediction given the current measurements, and $\mathcal{O}, \mathcal{U}$ are polyhedra that specify the variables constraints. At each time step the optimal inputs $u^*_k,\ldots,u^*_{k+N-1}$ are computed using QP, and only the first one is applied {to the system}, i.e. $u(k) = u^*_k$.

The plant dynamics used in the MPC solution are identified using the methodology summarized in Sec.~\ref{ssec:BackgroundSARX} and using a dataset consisting of 100 simulations, each one with time duration of 6 seconds, of the input and output of the plant with no packet losses. At any time $k$ we can use such model and the measurement of $x_k = x(k)$ to determine the switching sequence $i_0, \ldots, i_{N-1}$ and hence the matrices $A_{i_{j-1},i_{j}}, B_{i_{j-1},i_{j}}, F_{i_{j-1},i_{j}}$. The solution of Problem \ref{pbMPC-SA} also requires knowledge of the initial state of $\hat{\nu}_k$ and of its TPM $P$ \cite{Bernardini2011}: we derive two models of $\hat{\nu}_k$, respectively using the methodology in \cite{Lun2020} and in of Sec.~\ref{sec:RTMC}: for both the data-driven and physics-based channel Markov models, we set the number of channel operating modes equal to 9.

The cost function in Problem \ref{pbMPC-SA} models a tradeoff between penalizing deviations from the desired trajectory $x_k^*$ and minimizing the control effort. We define as control performance metric the cumulative cost of Problem \ref{pbMPC-SA} to compare the performance using the physics based channel model of Sec.~\ref{ssec:FSMC} as in \cite{Lun2020} and the data-driven channel model of  Sec.~\ref{sec:RTMC}.

	\begin{table}
		\centering
		\setlength{\extrarowheight}{5pt}
		\begin{tabular*}{0.95\columnwidth}{lcc}
			\toprule
			\textbf{Parameter}                                                  & \textbf{Value} & \textbf{Unit} \\ 
			\midrule
			Cart mass M                                                & 0.5           & kg       \\ 
			Pendulum mass m                                                & 0.2              & kg           \\  
			Distance from the pivot to the mass center L                             & 0.3              & m             \\ 
			Friction coefficient of the cart   d  & 0.1              & N$\cdot$s/m            \\
			Sampling time $T_u$					&0.001		&s \\
			\bottomrule
		\end{tabular*}
		\caption{Parameter values inverted pendulum}
		\captionsetup{justification=centering}
		\label{tab:plant}
	\end{table}
	
	\begin{table}
		\centering
		\setlength{\extrarowheight}{5pt}
		\begin{tabular*}{0.95\columnwidth}{lcc}
			\toprule
			\textbf{Parameter}                                                  & \textbf{Value} & \textbf{Unit} \\ 
			\midrule
			symbol rate 1/$T_s$                                                 & 62.5           & ksymb/s       \\ 
			channel bandwidth W                                                 & 2              & MHz           \\ 
			users speed  $v_0$, $v_1$                                           & 5.37           & m/s           \\ 
			shadowing decay distance $dc_0$, $dc_1$                             & 9              & m             \\ 
			shadowing standard dev.  $\sigma_{\beta_0}$, $\sigma_{\beta_1}$     & 2              & dB            \\
			power control error standard dev.  $\sigma_{e_0}$, $\sigma_{e_1}$   & 1.5            & dB            \\ 
			power control error decorr. time  $\tau_{\xi_0}$ and $\tau_{\xi_1}$   & 1.5            & dBm           \\ 
			reference user tx. power $P_0$                                      & 0              & dBm           \\ 
			interferer tx. power $P_1$                                          & 10             & dBm           \\ 
			distance reference tx-rx pair $d_1$                                 & 10             & m             \\ 
			\bottomrule
		\end{tabular*}
		\caption{Parameter values channel}
		\captionsetup{justification=centering}
		\label{tab:whart}
	\end{table}
	Tables \ref{tab:plant} and \ref{tab:whart} show the plant and channel parameters, respectively. The minimum update period $T_u = 0.1 s$ of the WirelessHART standard is too slow for several control applications, and makes the wireless link uncorrelated at the packet level. Thus, in view of showing the impact of our algorithms as a methodological enabler for the development of mobile network technologies that support much higher update rates, we consider $T_u$ = 0.001s. Moreover, to emphasise the impact and improvements of stochastic vs deterministic MPC, we set $\bar{d}_0 = 14$ to consider a scenario based on significant packet loss rates, where it is evident that deterministic MPC cannot guarantee acceptable performance while stochastic MPC does.

	\begin{figure}[htb]
		\centering
		\subfigure{\label{figState1}\includegraphics[width=0.9\columnwidth]{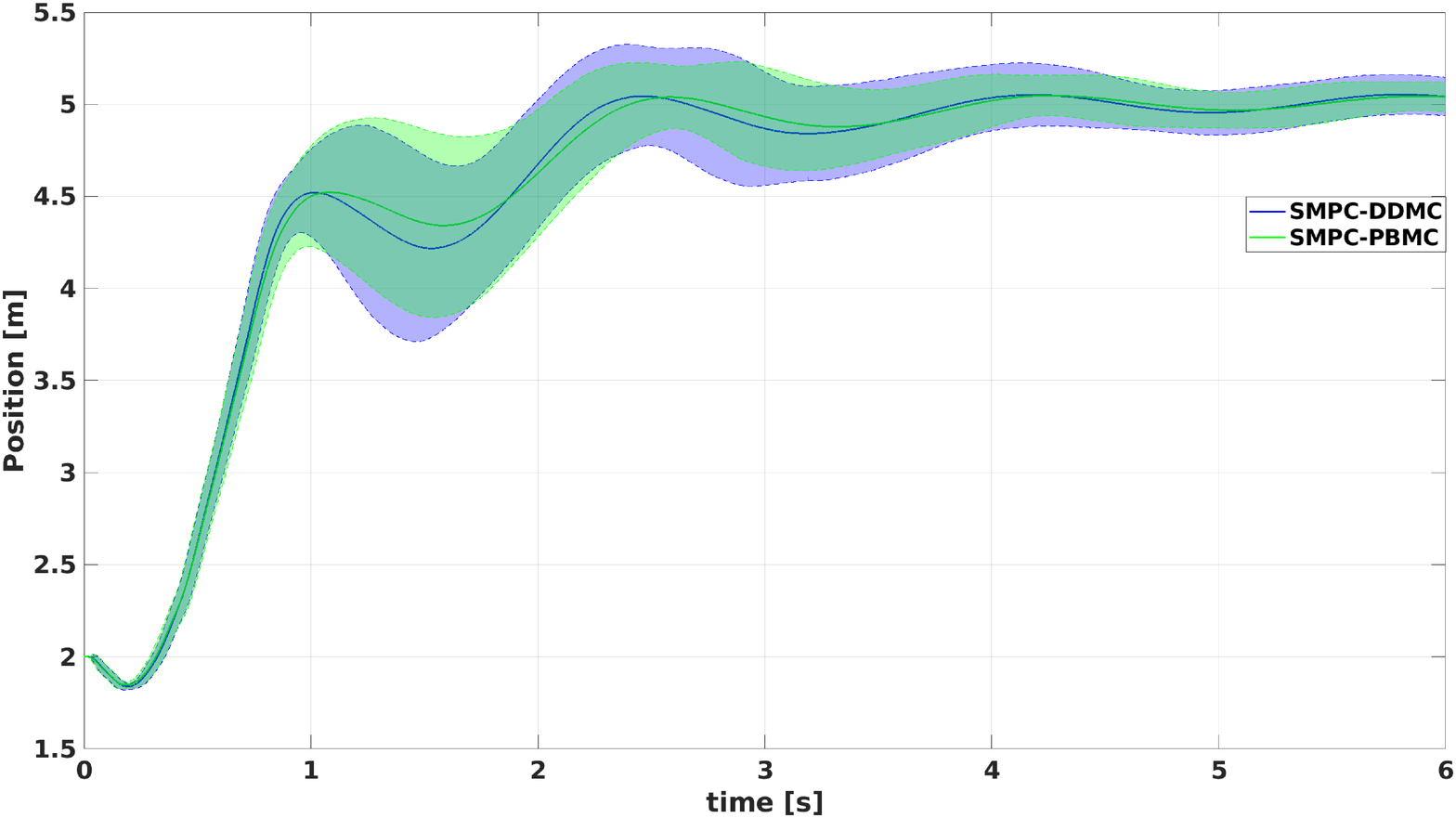}}
		\subfigure{\label{figState3}\includegraphics[width=0.9\columnwidth]{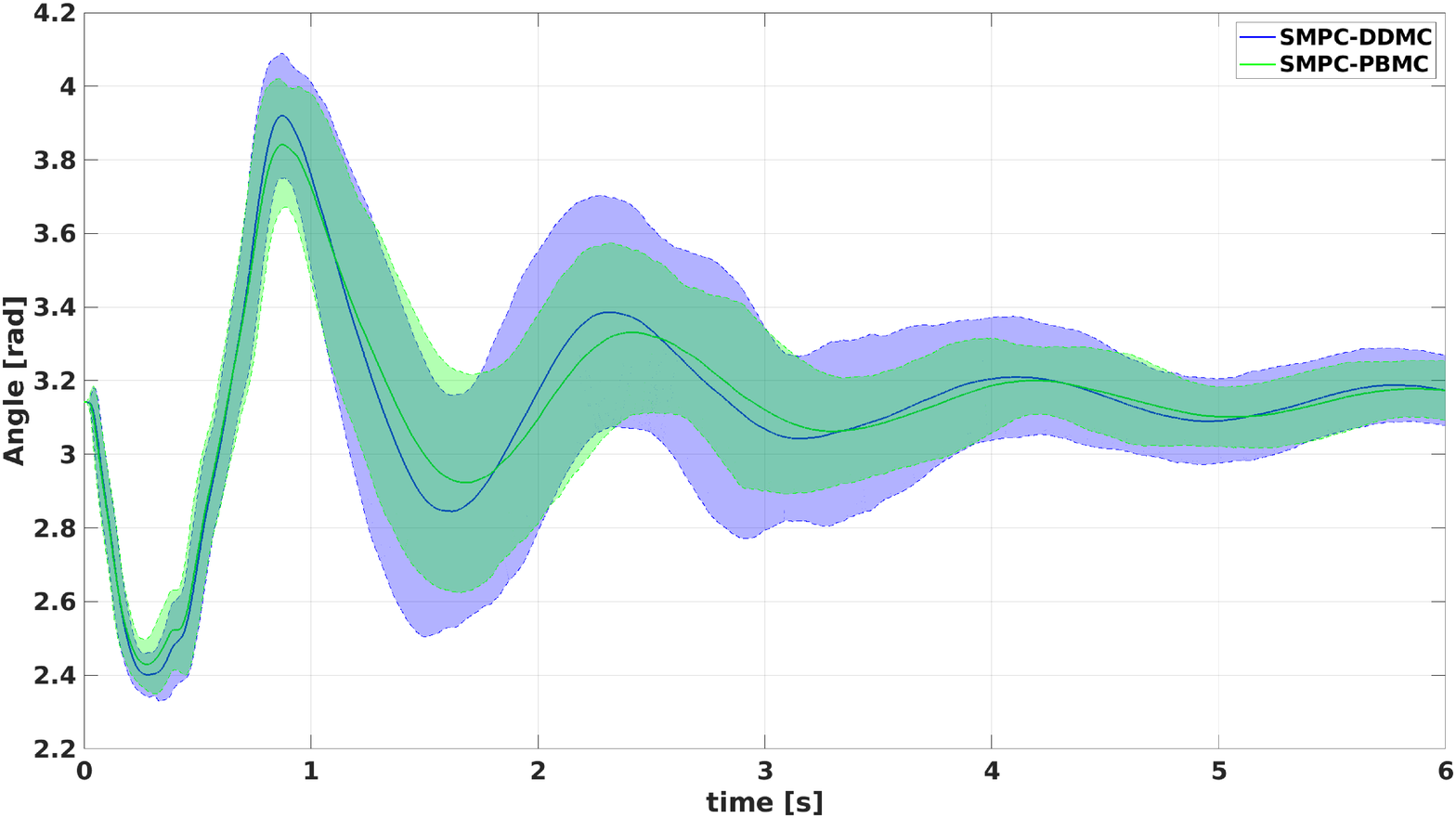}}
		\caption{Controlled states in the closed-loop simulation.}
		\label{figStates}
	\end{figure}
	
	\begin{figure}[h!]
		\begin{center}
			\centerline{\includegraphics[width=0.9\columnwidth]{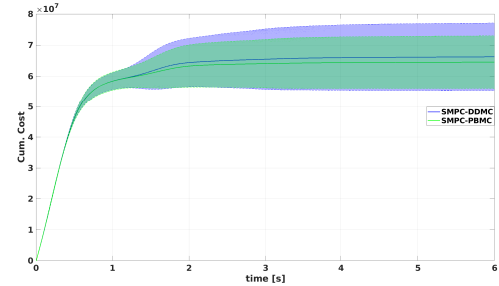}}
			\caption{Cumulative cost of the closed-loop simulation.}
			\label{figCumCost}
		\end{center}
	\end{figure}
	
	\begin{figure}[h!]
		\begin{center}
			\centerline{\includegraphics[width=0.9\columnwidth]{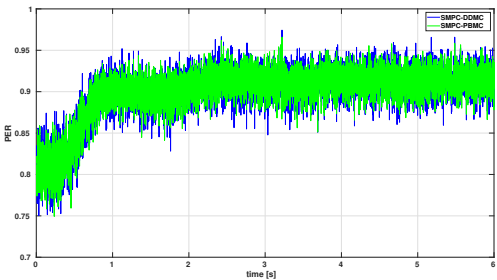}}
			\caption{PER of the closed-loop simulation.}
			\label{figPer}
		\end{center}
	\end{figure}
	To statistically validate the control performance we ran Monte Carlo simulations generating 500 admissible trajectories, each with $6000$ samples (corresponding to 6s). For all the simulations, the initial state is $x(0) = \left[2, 0, \pi, 0\right]^\top$ and the target state is $x^* = \left[5, 0, \pi, 0\right]^\top$. Figures \ref{figStates}, \ref{figCumCost} and \ref{figPer} illustrate simulation results. For each figure the averaged performance is displayed in solid line, and the $95.4\%$ confidence interval is represented with a shaded area. The performance of the controllers based on physics-based and data-driven models is very close, with the advantage of the data-driven approach that no a-priori knowledge of channel model and parameters is required.
	
	
	\section{Conclusions} \label{sec:Con}
	This paper provides a novel technique to learn Markov models representing fading wireless channels. We consider a validation scenario consisting of a WNCS that exploits a WirelessHART radio link to send the optimal control inputs generated by a Stochastic MPC, and show that the control performances of our data-driven approach and of a physics-based approach based on a stationary finite-state Markov chain are extremely close: this implies that in practical applications, when assuming perfect knowledge of the channel model and parameters is not possible, the methodology presented in this paper is a valid and very effective alternative. In future work we plan to validate our techniques in an experimental setup and to consider more general communication scenarios. 
	
	
	\bibliographystyle{IEEEtran}
	\bibliography{med_extended}

\begin{thebibliography}{10}
\providecommand{\url}[1]{#1}
\csname url@samestyle\endcsname
\providecommand{\newblock}{\relax}
\providecommand{\bibinfo}[2]{#2}
\providecommand{\BIBentrySTDinterwordspacing}{\spaceskip=0pt\relax}
\providecommand{\BIBentryALTinterwordstretchfactor}{4}
\providecommand{\BIBentryALTinterwordspacing}{\spaceskip=\fontdimen2\font plus
\BIBentryALTinterwordstretchfactor\fontdimen3\font minus
  \fontdimen4\font\relax}
\providecommand{\BIBforeignlanguage}[2]{{%
\expandafter\ifx\csname l@#1\endcsname\relax
\typeout{** WARNING: IEEEtran.bst: No hyphenation pattern has been}%
\typeout{** loaded for the language `#1'. Using the pattern for}%
\typeout{** the default language instead.}%
\else
\language=\csname l@#1\endcsname
\fi
#2}}
\providecommand{\BIBdecl}{\relax}
\BIBdecl

\bibitem{Park2018}
P.~{Park}, S.~{Coleri Ergen}, C.~{Fischione}, C.~{Lu}, and K.~H. {Johansson},
  ``Wireless network design for control systems: A survey,'' \emph{IEEE
  Communications Surveys Tutorials}, vol.~20, no.~2, pp. 978--1013, 2018.

\bibitem{Lu2016}
C.~{Lu}, A.~{Saifullah}, B.~{Li}, M.~{Sha}, H.~{Gonzalez}, D.~{Gunatilaka},
  C.~{Wu}, L.~{Nie}, and Y.~{Chen}, ``Real-time wireless sensor-actuator
  networks for industrial cyber-physical systems,'' \emph{Proceedings of the
  IEEE}, vol. 104, no.~5, pp. 1013--1024, 2016.

\bibitem{Alhen2019}
A.~{Ahlen}, J.~{Akerberg}, M.~{Eriksson}, A.~J. {Isaksson}, T.~{Iwaki}, K.~H.
  {Johansson}, S.~{Knorn}, T.~{Lindh}, and H.~{Sandberg}, ``Toward wireless
  control in industrial process automation: A case study at a paper mill,''
  \emph{IEEE Control Systems Magazine}, vol.~39, no.~5, pp. 36--57, 2019.

\bibitem{Chen2010}
A.~M. Deji~Chen, Mark~Nixon, \emph{WirelessHART}.\hskip 1em plus 0.5em minus
  0.4em\relax Springer, 2010.

\bibitem{Schenato2007}
L.~Schenato, B.~Sinopoli, M.~Franceschetti, K.~Poolla, and S.~S. Sastry,
  ``Foundations of control and estimation over lossy networks,''
  \emph{Proceedings of the IEEE}, vol.~95, no.~1, pp. 163--187, 2007.

\bibitem{Lun2019}
Y.~{Zacchia Lun} and A.~{D'Innocenzo}, ``Stabilizability of {Markov} jump
  linear systems modeling wireless networked control scenarios,'' in \emph{2019
  IEEE 58th Conference on Decision and Control (CDC)}, 2019, pp. 5766--5772.

\bibitem{Turin1998}
W.~{Turin} and R.~{van Nobelen}, ``Hidden markov modeling of flat fading
  channels,'' \emph{IEEE Journal on Selected Areas in Communications}, vol.~16,
  no.~9, pp. 1809--1817, 1998.

\bibitem{Bilmes1998b}
J.~A. Bilmes, ``A gentle tutorial of the em algorithm and its application to
  parameter estimation for gaussian mixture and hidden markov models (no.
  tr-97-021),'' \emph{Berkeley, CA: International Computer Science Institute
  and Univ. Calif. Berkeley}, 1998.

\bibitem{Lawrance1989}
R.~Lawrance and A.~Rabiner, ``A tutorial on hidden markov models and selected
  applications in speech recognition,'' \emph{Proceedings of the IEEE},
  vol.~77, no.~2, pp. 257--286, 1989.

\bibitem{Salamatian2001}
\BIBentryALTinterwordspacing
K.~Salamatian and S.~Vaton, ``Hidden markov modeling for network communication
  channels,'' \emph{SIGMETRICS Perform. Eval. Rev.}, vol.~29, no.~1, p.
  92–101, Jun. 2001. [Online]. Available:
  \url{https://doi.org/10.1145/384268.378439}
\BIBentrySTDinterwordspacing

\bibitem{Sadghi2008}
P.~Sadeghi, R.~Kennedy, P.~Rapajic, and R.~Shams, ``Finite-state markov
  modeling of fading channels,'' \emph{IEEE Signal Processing Magazine},
  vol.~57, 2008.

\bibitem{Lun2020}
Y.~{Zacchia Lun}, C.~Rinaldi, A.~Alrish, A.~{D'Innocenzo}, and F.~Santucci,
  ``On the impact of accurate radio link modeling on the performance of
  wirelesshart control networks,'' in \emph{IEEE INFOCOM 2020-IEEE Conference
  on Computer Communications}.\hskip 1em plus 0.5em minus 0.4em\relax IEEE,
  2020, pp. 2430--2439.

\bibitem{Breiman2017classification}
L.~Breiman, \emph{Classification and regression trees}.\hskip 1em plus 0.5em
  minus 0.4em\relax Routledge, 2017.

\bibitem{SmarraNAHS2020}
F.~Smarra, G.~D. Di~Girolamo, V.~De~Iuliis, A.~Jain, R.~Mangharam, and
  A.~{D'Innocenzo}, ``Data-driven switching modeling for mpc using regression
  trees and random forests,'' \emph{Nonlinear Analysis: Hybrid Systems},
  vol.~36, p. 100882, 2020.

\bibitem{2006}
{IEEE 802.15.4-2006}, ``{{IEEE} Standard for Information technology -- Local
  and metropolitan area networks -- Specific requirements -- Part 15.4:
  Wireless Medium Access Control ({MAC}) and Physical Layer ({PHY})
  Specifications for Low Rate Wireless Personal Area Networks ({WPANs})},''
  IEEE, Standard, September 2006.

\bibitem{Goldsmith2005}
A.~Goldsmith, \emph{Wireless communications}.\hskip 1em plus 0.5em minus
  0.4em\relax Cambridge university press, 2005.

\bibitem{fischione2007}
C.~Fischione, F.~Graziosi, and F.~Santucci, ``Approximation for a sum of
  {On-Off} log-normal processes with wireless applications,'' \emph{IEEE Trans.
  Commun.}, vol.~55, no.~9, pp. 1822--1822, Sept. 2007.

\bibitem{Baddour2005}
K.~E. {Baddour} and N.~C. {Beaulieu}, ``Autoregressive modeling for fading
  channel simulation,'' \emph{IEEE Transactions on Wireless Communications},
  vol.~4, no.~4, pp. 1650--1662, Jul. 2005.

\bibitem{Kay1988}
S.~M. Kay, \emph{Modern spectral estimation}.\hskip 1em plus 0.5em minus
  0.4em\relax Pearson Education India, 1988.

\bibitem{Bernardini2011}
D.~Bernardini and A.~Bemporad, ``Stabilizing model predictive control of
  stochastic constrained linear systems,'' \emph{IEEE Transactions on Automatic
  Control}, vol.~57, no.~6, pp. 1468--1480, 2011.

\end{thebibliography}
\end{document}